\documentclass[11pt]{llncs}
\usepackage{listings}
\usepackage{graphicx}
\usepackage{url}
\begin{document}


\lstset{language=C++}
\title{Buffer overflow vulnerabilities in CUDA: a preliminary analysis}

\author{Andrea Miele}
\institute{EPFL, LACAL, Lausanne, Switzerland}

\maketitle

\subsection*{Abstract}
We present a preliminary study of buffer overflow vulnerabilities in CUDA software running on GPUs.
We show how an attacker can overrun a buffer to corrupt sensitive data or steer the execution flow by overwriting function pointers, e.g., manipulating the virtual table of a C++ object.
In view of a potential mass market diffusion of GPU accelerated software this may be a major concern.
\section{Introduction}
General-Purpose Computing on Graphics Processing Units (GPGPU) has become very popular in recent years.
Modern GPUs are many-core accelerators for massively parallel applications rather than simply 2D and 3D graphics rendering coprocessors as in the past.
Hardware/software platforms for GPGPU like Compute Unified Device Architecture (CUDA)~\cite{CUDAProgramming2014} and OpenCL~\cite{opencl,OpenCLAMD} allow developers to program graphics cards in languages like C/C++ or Fortran through APIs and language extensions.
Such platforms have been used to accelerate a variety of scientific applications in the recent past and their use to speed-up commodity applications (e.g., web applications~\cite{webgl}) or operating systems can be expected in the near future~\cite{gpufs}.
As a consequence, questions about the security of code running on these platforms have become relevant.
So far the security of GPU code has been somehow touched upon only considering GPUs as accelerators for malware~\cite{5665801,Ruxcon} and more recently information leakage through side channels~\cite{cudaleaks}, but as far as we know no one has explored potential vulnerabilities of GPU code and their exploitation.
A natural question that arises is whether or not classic C/C++ ``buffer'' overflow vulnerabilities~\cite{bof,hof,vtabo} can be exploited. 
Leveraging a vulnerability to run arbitrary code on CPUs usually aims at interacting with the operating system via system calls to perform privilege escalation.
In the context of GPUs there is no notion of operating system and the code runs on many cores across a large number of threads.
Thus, the concept of exploit assumes a different dimension.
The ultimate goal of an attacker may be tampering with a parallel computation to maliciously affect the outcome or to force one or more threads to jump to specific parts of the code.
This may become critical in view of future tight integration of CPU and GPU architectures.

The study of this problem on current GPUs is hindered by the lack of documentation on low level details of the hardware/software interfaces.
The actual ISA of GPUs is usually not exposed by vendors who provide, in some cases, only virtual low level assembly~\cite{PTX}.
Recently NVIDIA has released a disassembler for their binary executable format and a list of ISA instructions~\cite{cudabin}.
The latter, though, does not include a description of their syntax and functioning.
Consequently, the analysis of GPU software vulnerabilities has to rely on ``trial and error'' experiments and reverse engineering.

In this paper we present a preliminary study of buffer overflow vulnerabilities in CUDA software.
In particular we show how a buffer overrun can be exploited to overwrite function pointers (e.g., to manipulate the virtual table of a C++ object) to steer the execution flow.
An attacker can hijack each call to specific functions to other functions in the code of his/her choice or perform limited return orient programming (ROP)~\cite{rop}.
We hope that our results will act as a wake-up call for the community as we believe that GPU software security will soon become an extremely relevant problem.
The source code of this project will be made freely available.
\section{Compute Unified Device Architecture (CUDA)}
\label{sec:CUDA}
In this section we give a brief overview of CUDA.
CUDA~\cite{CUDAProgramming2014} is a computing platform, consisting in both a hardware and a software architecture, enabling NVIDA GPUs to support general purpose computing. 
At the programming level CUDA consists of extensions to the C/C++ language, libraries and some specific data types that enable the programmer to compute on the GPU.
An actual function call mechanism exists and defining recursive functions is possible. Moreover, OOP programming is possible through a subset of C++ constructs that are supported. 
In CUDA programs the code that runs on the GPU is enclosed in a special function called \textit{kernel}. 
A kernel is executed in the form of multiple parallel instances corresponding to a set of parallel \textit{threads}. 
Threads are grouped in \textit{blocks} and blocks are grouped in \textit{grids}:
\begin{itemize}
\item \textbf{thread:} A thread executes one instance of the kernel, and it is uniquely identified inside its block by a thread identifier. 
Each thread has its program counter, registers, per-thread private memory, input, and output results. 
\item \textbf{block:} A block is a set of concurrently executing threads that can cooperate among themselves through synchronization and shared memory.
Each thread block has a private per-block shared memory space used for inter-thread communication, data sharing, and result sharing in parallel algorithms.
\item \textbf{grid:} A grid is an array of thread blocks that execute the same kernel, read inputs from global memory, write results to global memory, and synchronize between dependent kernel calls.
The GPU executes a kernel as a grid of parallel thread blocks.
\end{itemize}
This hierarchal grouping scheme allows CUDA applications to scale across different device models.
See~\cite{CUDAProgramming2014} for more details.

\section{Practical analysis}
In this section we present our practical analysis through two examples.
We will consider the presently common CPU/GPU interaction model in which the CPU and GPUs work on separate physical memory.
The CPU loads the input data into the GPU memory and then runs the GPU kernel that will process these data and produce output data.
A soon as the kernel terminates, the CPU copies the produced output from the GPU memory to the CPU memory.
We will not explore the case where CPU and GPU are somehow integrated and share the same memory.

In our first example we show how function pointers in static memory can be overwritten to have each thread call a function that should normally not be callable.
In our second example we show how the VTABLE of a C++ object in dynamic memory can be manipulated for the same purpose.
Table \ref{tab:specs} shows the specifications of the platform we have used for our experiments.
\begin{table}[!htp]
\centering
\caption{Experimental platform specifications.}
\label{tab:specs}
\begin{tabular}{|l|l|}
\hline
CPU & Intel Core i7-4790, 3.60GHz \\ \hline
GPU & Asus GTX Titan Black, 2880 cores\\ \hline
OS & Ubuntu 14.04.1 LTS 64-bit \\ \hline
Compiler & CUDA nvcc 6.5 \\ \hline
\end{tabular}
\end{table}
\subsection{A stack overflow}
\label{stack}
Consider the following fragment of CUDA code (the qualifier \texttt{\_\_device\_\_} denotes a function that is called by threads running on the GPU).
{\tt \small
\begin{verbatim}
#define BUF_LEN 16

typedef unsigned long(*pFdummy)(void);

__device__ __noinline__ unsigned long dummy1(){
 return 0x1111111111111111;
}
__device__ __noinline__ unsigned long dummy2(){
 return 0x2222222222222222;
}
__device__ __noinline__ unsigned long dummy3(){
 return 0x3333333333333333;
}
__device__ __noinline__ unsigned long dummy4(){
 return 0x4444444444444444;
}
__device__ __noinline__ unsigned long dummy5(){
 return 0x5555555555555555;
}
__device__ __noinline__ unsigned long dummy6(){
 return 0x6666666666666666;
}
__device__ __noinline__ unsigned long dummy7(){
 return 0x7777777777777777;
}
__device__ __noinline__ unsigned long dummy8(){
 return 0x8888888888888888;
}
__device__ __noinline__ unsigned long dummy9(){
 printf("HELLO ADMIN!\n");
 return 0x9999999999999999;
}

__device__ __noinline__ unsigned long unsafe(
unsigned int * input, int len){

 unsigned int buf[BUF_LEN];
 pFdummy fp[8];
 fp[0]=dummy1;
 fp[1]=dummy2;
 fp[2]=dummy3;
 fp[3]=dummy4;
 fp[4]=dummy5;
 fp[5]=dummy6;
 fp[6]=dummy7;
 fp[7]=dummy8;
 unsigned int hash = 5381;

 // copy input to buf
 for(int i=0;i<len;i++)
  buf[i]=input[i];
 //djb2
 for(int i=0;i<BUF_LEN;i++){
  hash = ((hash << 5) + hash) + buf[i];
 }
 return (unsigned long) (fp[hash%8])();
}
\end{verbatim}
}
The function \texttt{unsafe} is vulnerable to a buffer overflow (the array \texttt{buf} can be overridden if the attacker has control over the variables \texttt{string} and \texttt{len}).
It computes a hash value of the input (using the djb2 algorithm of D.J. Bernstein) and uses such value reduced modulo 8 to select and call one of the first 8 ``dummy'' functions.
The code of the simple CUDA kernel using the \texttt{unsafe} functions is the following:
{\tt \small
\begin{verbatim}
__global__ void test_kernel(unsigned long* hashes,
unsigned int * input, int len, int *admin){
 unsigned long my_hash;
  int idx=blockDim.x*blockIdx.x+threadIdx.x;
 if(*admin)
  my_hash=dummy9();
 else
  my_hash=unsafe_hash(input+(len*idx),len);
 hashes[idx]=my_hash;
} 
\end{verbatim}
}
It is possible to use \texttt{cuda-gdb} or disassemble the binary with \texttt{cuobjdump} to figure out what are the addresses of the dummy functions. For instance on our platform the function \texttt{dummy9} has address~ \texttt{0x4e0}.
This address is relative to the base address of the code section and does not change across multiple executions.
We launch our kernel on one thread with value pointed by \texttt{admin} set to zero (we omit the very simple host code for the sake of clarity) and observe that if we fill the input buffer with at most 26 values (the value of \texttt{len} is set to 26), for instance 26 times \texttt{0x4e0}, the code prints correctly: \texttt{Hash[0]: 6666666666666666}.
If we fill the input buffer with one more value \texttt{0x4e0} (the address of~\texttt{dummy9}) and set the value of \texttt{len} to 27, the output is instead \texttt{HELLO ADMIN! Hash[0]: 9999999999999999}, thus we have successfully overwritten the function pointers with the address of~\texttt{dummy9}.
If more than one thread is run we observe that the each thread is hijacked to execute~\texttt{dummy9}.
By looking at the disassembled binary with \texttt{cuobjdump}: {\tt \small
\begin{verbatim}
     .......................................
 /*03d8*/  LDL R0, [R0];          
 /*03e0*/  PRET 0x3f0;          
 /*03e8*/  BRX R0 -0x3f0;        
 /*03f0*/  IADD R1, R1, 0x80;     
 /*03f8*/  RET;                 
                             
 /*0408*/  MOV32I R4, 0x11111111; 
 /*0410*/  MOV32I R5, 0x11111111; 
 /*0418*/  RET;                
 /*0420*/  MOV32I R4, 0x22222222
 /*0428*/  MOV32I R5, 0x22222222
 /*0430*/  RET;                
 /*0438*/  MOV32I R4, 0x33333333
                             
 /*0448*/  MOV32I R5, 0x33333333
 /*0450*/  RET;                
 /*0458*/  MOV32I R4, 0x44444444
 /*0460*/  MOV32I R5, 0x44444444
 /*0468*/  RET;                
 /*0470*/  MOV32I R4, 0x55555555
 /*0478*/  MOV32I R5, 0x55555555
                             
 /*0488*/  RET;                
 /*0490*/  MOV32I R4, 0x66666666
 /*0498*/  MOV32I R5, 0x66666666
 /*04a0*/  RET;                
 /*04a8*/  MOV32I R4, 0x77777777
 /*04b0*/  MOV32I R5, 0x77777777
 /*04b8*/  RET;                
                             
 /*04c8*/  MOV32I R4, 0x88888888
 /*04d0*/  MOV32I R5, 0x88888888
 /*04d8*/  RET;                
 /*04e0*/  MOV32I R4, 0x0;       
 /*04e8*/  MOV32I R5, 0x0;       
 /*04f0*/  MOV R7, RZ;          
 /*04f8*/  MOV R6, RZ;          
                             
 /*0508*/  JCAL 0x0;            
 /*0510*/  MOV32I R4, 0x99999999
 /*0518*/  MOV32I R5, 0x99999999
 /*0520*/  RET;                
 /*0528*/  BRA 0x528;           
 /*0530*/  NOP;                
 /*0538*/  NOP;                
     .......................................
\end{verbatim}
}
we observe that the address \texttt{0x4e0} of \texttt{dummy9} is relative to the base address of the code as mentioned above. 
The ``Branch to Relative Indexed Address'' \texttt{BRX} instruction at address~\texttt{0x03e8} is used to jump to the correct dummy function, whose address is stored in register~\texttt{R0}.

It follows that by overwriting the function pointers in \texttt{fp} an attacker can jump to any address in the code memory and so ROP type of exploits are theoretically possible.
Notice that the function call and return mechanism is handled with a \texttt{PRET} instruction followed eventually by a \texttt{RET} instruction. The \texttt{PRET} instruction presumably stores the return address (again relative to the code base address) at an unknown location.
Classic return address overwrite attacks seem not to be feasible and attempts to jump outside the code memory failed (for instance we could not find a way to jump to shellcode injected into the buffer~\texttt{buf}).
Therefore, we conclude that code and data address spaces are separated and thus simple injected code execution is not possible.
\subsection{A ``heap overflow'': manipulating the virtual table of a C++ object}
Consider the following code where we define a C++ class B with four virtual methods: \texttt{f1}, \texttt{f2}, \texttt{f3} and~\texttt{f4} and
the derived class D defines the above four methods.
The function \texttt{unsafe} is very similar to the homonymous function described in section~\ref{stack}, but in this case the array \texttt{buf} and a class D object are dynamically allocated.
The function is vulnerable to a ``heap'' overflow as the array \texttt{buf} can be overridden to overwrite the adjacent class D object. 
{\tt \small
\begin{verbatim}

#define BUF_LEN 8

class B    //base class
{
  public:
    __device__  virtual unsigned long f1
    (unsigned int hash)
     {return 0;}
    __device__  virtual unsigned long f2
    (unsigned int hash)
     {return 0;}
    __device__  virtual unsigned long f3
    (unsigned int hash) 
    {return 0;}
    __device__  virtual unsigned long f4
    (unsigned int hash) 
    {return 0;}
};

class D : public B { 
  public:
    __device__ __noinline__ unsigned long f1
    (unsigned int hash);
    __device__ __noinline__ unsigned long f2
    (unsigned int hash);
    __device__ __noinline__ unsigned long f3
    (unsigned int hash);
    __device__ __noinline__ unsigned long f4
    (unsigned int hash);
};
__device__ __noinline__ unsigned long D::f1
(unsigned int hash) 
{return hash;}
__device__ __noinline__ unsigned long D::f2
(unsigned int hash) 
{return 2*hash;}
__device__ __noinline__ unsigned long D::f3
(unsigned int hash) 
{return 3*hash;}
__device__ __noinline__ unsigned long D::f4
(unsigned int hash) 
{return 4*hash;}

__device__ __noinline__ unsigned long secret() {
  printf("HELLO ADMIN! ");
  return 0x9999999999999999;
};

__device__ __noinline__ unsigned long unsafe(
unsigned long * input, unsigned int len){
  unsigned long res =0;
  unsigned long hash = 5381;
  unsigned long *buf =
  (unsigned long *) 
  malloc(sizeof(unsigned long)*BUF_LEN);
  D *objD = new D;

  // copy input to buf
  for(int i=0;i<len;i++)
    buf[i]=input[i];
  //djb2
  for(int i=0;i<BUF_LEN;i++){
    hash = ((hash << 5) + hash) +buf[i];
  }

  res=objD->f1(hash);
  res=objD->f2(res);
  res=objD->f3(res);
  res=objD->f4(res);
  free(buf);
  delete objD;
  return res;
}
\end{verbatim}
}
The kernel we focus on is also very similar to the kernel presented in section~\ref{stack}:
{\tt \small
\begin{verbatim}
__global__ void test_kernel(unsigned long* hashes,
unsigned long * input, unsigned int len, int *admin){
  unsigned long my_hash;
  int idx=blockDim.x*blockIdx.x+threadIdx.x;
  if(*admin)
    my_hash=secret();
  else
    my_hash=unsafe_hash(input+(len*(idx)),len);
  hashes[idx]=my_hash;
}
\end{verbatim}
}
By instrumenting the code with simple \texttt{printf} calls we have been able to observe the following:
\begin{itemize}
\item The addresses of dynamically allocated memory blocks (\texttt{malloc}) or objects (\texttt{new}) are predictable.
In our case, considering the first thread of the first block, the address of the array \texttt{buf} allocated in the function \texttt{unsafe} is always~\texttt{0xb0513f920}.
\item The first 64-bit value stored in an object of class D is the address of its virtual table (VTABLE). This address is the same across different threads as one would expect.
\item  The VTABLE contains the addresses (relative to the base of code section) of the four virtual functions defined in class D. This can be verified printing the VTABLE and comparing the first four 64-bit values with the addresses
of the four functions appearing in the disassembled code obtained with~\texttt{cuobjdump}.
\end{itemize}
Using the above observations we can exploit the overflow vulnerability to overwrite the VTABLE address of the class D object
instantiated in a thread with the address of a ``forged'' VTABLE (which is stored in \texttt{buf}) containing 4 copies of the address of the function~\texttt{secret}.
Then the thread is forced to call the function~\texttt{secret}.

\begin{figure}[!ht]
\centering
\includegraphics[width=1\columnwidth]{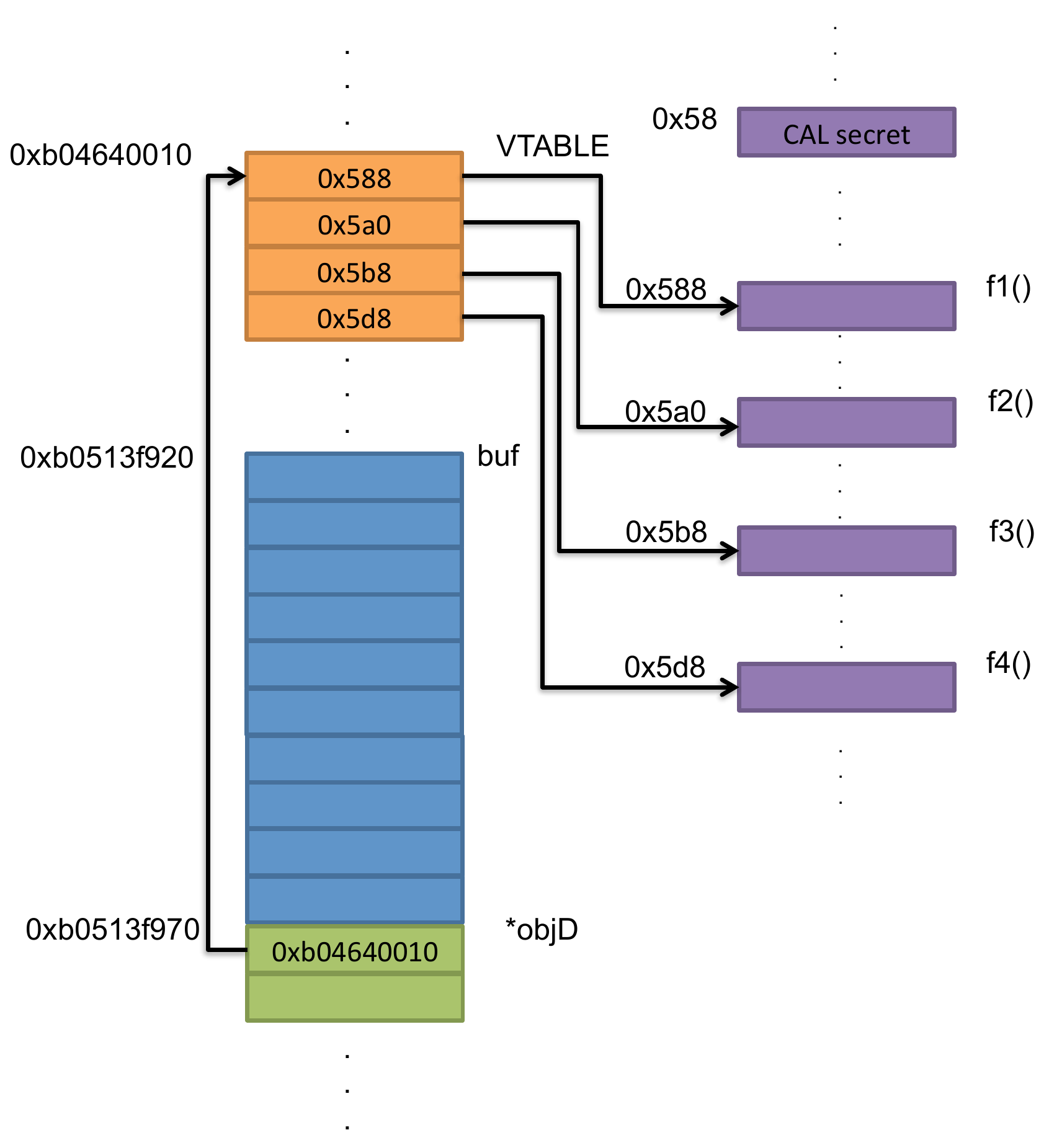}
\caption{Memory layout snapshot for function~\texttt{unsafe}.}\label{fig:normal}
\end{figure}

\begin{figure}[!ht]
\centering
\includegraphics[width=1\columnwidth]{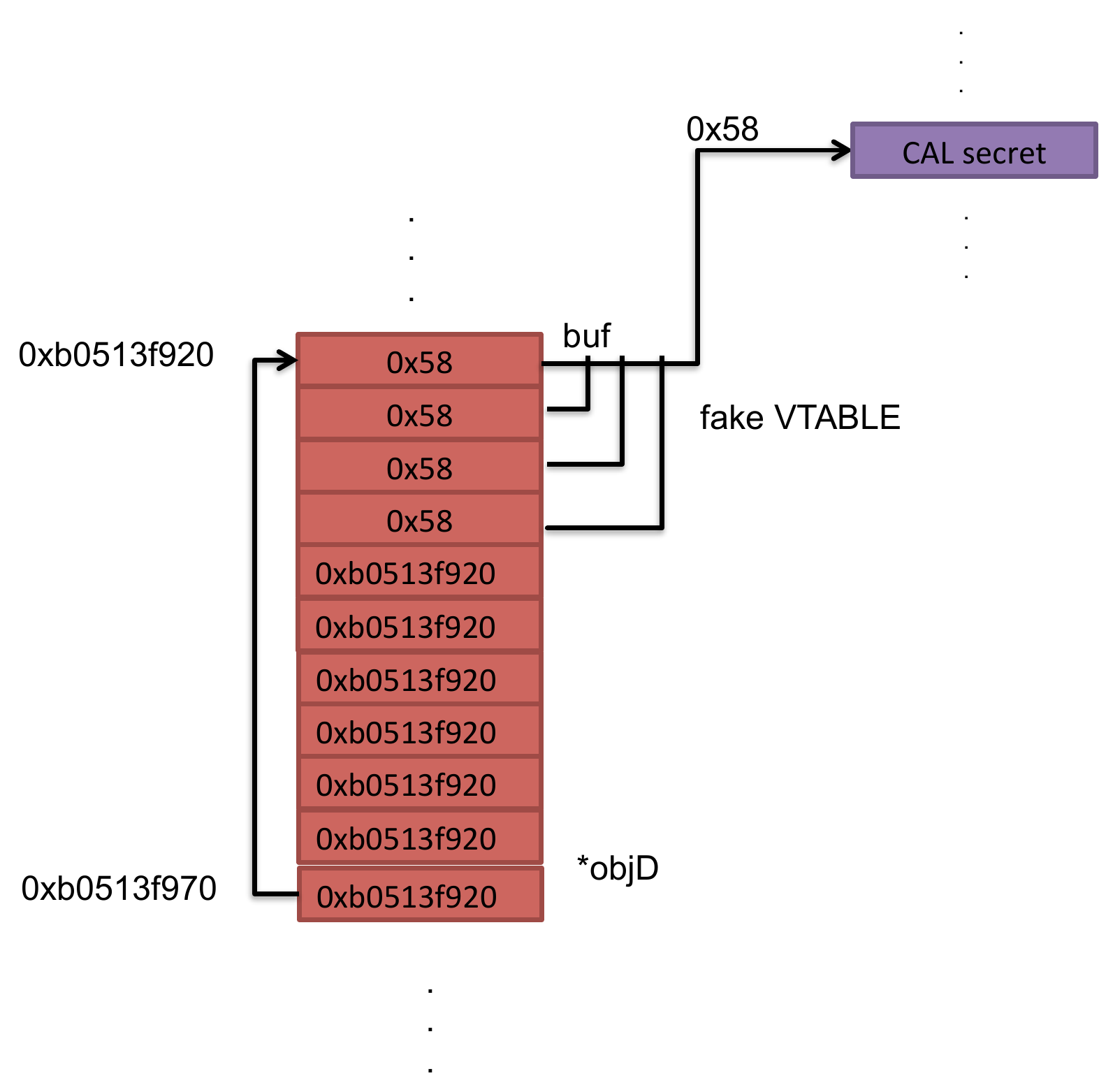}
\caption{Exploiting the heap overflow in function~\texttt{unsafe}. We override the buffer~\texttt{buf} to inject a forged VTABLE into the object~\texttt{*objD}.}\label{fig:attack}
\end{figure}

Focusing on a single thread we have the memory layout showed in Figure~\ref{fig:normal}.   
The exploit is achieved by filling the array \texttt{buf} as depicted in Figure~\ref{fig:attack}, namely the first 11 locations of \texttt{buf} must contain the address of the the forged VTABLE, which is in turn stored in  
\texttt{buf} starting at the 12-th location.
We have not discovered any hint suggesting the use of classic \texttt{malloc} linked lists with pointers stored next to the actual data.
We leave the exploration of classic \texttt{malloc} and \texttt{free} exploits~\cite{hof} as future work.
\section{Conclusion}
We have shown that CUDA software running on the latest NVIDIA GPUs can be vulnerable to classic buffer overflow attacks.
Buffer overflows in both static and dynamic memory can be exploited to overwrite sensitive data or function pointers (e.g., to manipulate a C++ object virtual table).
Our analysis does not expose any concrete threat, however it shows that the exploitation of vulnerabilities in GPU software is possible.
This may become a critical problem in the future if commodity GPGPU software will spread, especially if GPUs and CPUs will be tightly integrated.

\bibliographystyle{abbrv}
\bibliography{bib}


\end{document}